\documentclass[12pt]{article}
\usepackage{epsf}
\title{{\bf Small size pentaquark width: calculation in QCD sum rules}}

\author{A.G.Oganesian\\
Institute of Theoretical and Experimental Physics,\\
B.Cheremushkinskaya 25, 117218 Moscow,Russia}
\date{}

\begin{document}

\maketitle

\newcommand{\be}{\begin{equation}}
\newcommand{\ee}{\end{equation}}

\def\la{\mathrel{\mathpalette\fun <}}
\def\ga{\mathrel{\mathpalette\fun >}}
\def\fun#1#2{\lower3.6pt\vbox{\baselineskip0pt\lineskip.9pt
\ialign{$\mathsurround=0pt#1\hfil##\hfil$\crcr#2\crcr\sim\crcr}}}

\vspace{1cm}
\begin{abstract}

The pentaquark width is calculated in QCD sum rules. The higher
dimension operators contribution is accounted. It is shown, that
$\Gamma_{\Theta}$ should be very small, less than $1Mev$.

\end{abstract}


\vspace{1cm}

\normalsize

The status  of $\Theta^+$, predicted in 1997 by D.Diakonov,
V.Petrov and M.Polyakov \cite{n4} in the Chiral Soliton Model,
till now is doubtful. Few years ago narrow exotic baryon resonance
$\Theta^+$ with quark content $\Theta^+ = uudd\bar{s}$ and mass
1.54 GeV had been discovered by two groups \cite{n1,n2}. But the
question is open until now, during last two years some groups
confirm the  pentaquark $\Theta^+$ existence, while other see null
signal. Moreover, last year some groups, which have seen
pentaquark, in the new experiments with higher statistics reported
null result for pentaquark signal (CLAS experiments on hydrogen
and deuterium $\cite{ex1}$, BELLE $\cite{ex2}$) but at the same
time DIANA $\cite{dolg}$, and also LEPS,  SVD-2 confirm their
results with higher statistic(see $\cite{buk}$ for the review).
(Some theoretical reviews are given in \cite{n5,n6}).

So yet one can say only that if pentaquark exist, it should be a
narrow state. Experimentally, only an upper limit was found, the
stringer bound was presented in \cite{n2}: $\Gamma < 9 MeV$. The
phase analysis of $KN$ scattering results in the even stronger
limit on $\Gamma$ \cite{n7}, $\Gamma < 1 MeV$. A close to the
latter limitation was found in \cite{n8} from the analysis of $Kd
\to ppK$ reaction and in \cite{n9} from $K+Xe$ collisions data
\cite{n2}. Also $\cite{ex2}$ from the negative result of the
experiment give the upper limit for pentaquark width less than
640$KeV$

In the paper \cite{ja1}, \cite{ja2} it was shown, that if
pentaquark is genuine state it width should be strongly
parametrically suppressed. It is necessary to note that this is
general statement and   does not depend of the choice of the
pentaquark current (without derivatives), but at the same time it
is significantly based on the assumption, that the size of the
pentaquark is not larger, than usual hadronic. The main goal of
this paper is to find numerical estimation of pentaquark width by
use the method offered in \cite{ja1},\cite{ja2}.

 \bigskip

{\bf \large Part 1. } In the papers \cite{ja1}, \cite{ja2} it was
shown, that pentaquark width should be suppressed as
$\Gamma_{\Theta} \sim \alpha^2_s \langle 0 \vert \bar{q} q \vert 0
\rangle^2$,  (for any current without derivatives). Later, in the
short paper \cite{ja4} the first non-vanishing operator (dimension
$d=3$) contribution was calculated and the sum rule was considered
numerically. It was shown that pentaquark width is suppressed
numerically also and the width of the pentaquark width was
estimated to be less 1 Mev. In this paper we will discuss this sum
rules in more detail and also the contribution of the operators of
the higher dimensions will be accounted. Let us shortly remind the
main points of the method. We start from 3-point correlator

\be
 \Pi_{\mu}=\int e^{i(p_1x-qy)} \langle 0\mid
\eta_{\theta}(x)j^5_{\mu}(y) \eta_n(0)\mid 0 \rangle \ee

where $\eta_n(x)$ is the neutron quark current \cite{nbl1},
($\eta_n =\varepsilon^{abc} (d^a C \gamma_{\mu} d^b) \gamma_5
\gamma_{\mu} u^c $),

$ \langle 0\mid \eta_n\mid n\rangle =\lambda_n v_n$,
($v_n$ is the  nucleon spinor), $\eta_{\theta}$ is an arbitrary
pentaquark current
 $ \langle 0\mid \eta_{\theta}\mid \theta^+
\rangle=\lambda_{\theta} v_{\theta}$ and $j_{\mu 5} = \bar{s}
\gamma_{\mu}\gamma_5 u$ is the strange axial current.

As an example of $\eta_{\theta}$ one can use the following one
(see \cite{jap}, where it was first offered, and also \cite {ja3},
where the sum 2-point rule analysis for this current was
discussed):

\be
J_A =\varepsilon^{abc} \varepsilon^{def} \varepsilon^{gcf}
(u^{a^T} Cd^b ) (u^{d^T} C\gamma^{\mu} \gamma_5 d^e)\gamma^{\mu}
c\bar{s}_g\ee

and we will use it farther to obtain numerical results. (Of course
there is large number of the another currents, for example see
\cite {koch}, where 2-point correlators was analyzed very
carefully, taking into account operators up to dimension 13 and
direct instanton contribution).

As usual in QCD sum rule the physical representation of correlator
(1) can be saturated by lower resonance states plus continuum
(both in $\eta_{\Theta}$ and nucleon channel)
\be
 \Pi^{Phys}_{\mu}=\langle 0\mid \eta_{\theta}\mid \theta^+ \rangle
\langle \theta^+ \mid j_{\mu}\mid n \rangle \langle n\mid \eta_n
\mid 0 \rangle \frac{1}{p^2_1 -m^2_{\theta}}\frac{1}{p^2_2-m^2}+
cont. \ee

where $p_2=p_1-q$ is nucleon momentum, $m$ and
$m_{\theta}$ are nucleon and pentaquark masses.

Obviously, in the limit of massless kaon \be \langle \theta^+ \mid
j_{\mu}\mid n \rangle =g^A_{\theta n} \bar{v}_n \Biggl (g^{\mu\nu}
-\frac{q^{\mu}q^{\nu}}{q^2}\Biggr ) \gamma^v\gamma_5 v_{\theta}
\ee

where axial transition constant $g^A_{\theta n}$ is just we are
interesting in (the width is proportional to the square of this
value). Such a method for calculation the width in QCD sum rules
is not new, see, e.g. \cite{nblk0}. In the case of massive kaon
the only change is in denominator of second term in r.h.s of the
eq. (4), i.e. \be \langle \theta^+ \mid j_{\mu}\mid n \rangle
=g^A_{\theta n} \bar{v}_n \Biggl (g^{\mu\nu}
-\frac{q^{\mu}q^{\nu}}{q^2-m_k^2}\Biggr ) \gamma^{\nu}\gamma_5
v_{\theta} \ee It is clear that the second term vanishes at small
$q^2$.

Substituting  $ \langle 0\mid \eta_n\mid n\rangle =\lambda_n v_n$,
 and  $ \langle 0\mid \eta_{\theta}\mid \theta^+
\rangle=\lambda_{\theta} v_{\theta}$  in eq (3) and take the sum
on polarization one can easily see, that (in the limit of small
$q^2$) correlator (1) is proportional to $g^A_{\theta n}$.

\be
 \Pi^{Phys}_{\mu}=\lambda_n\lambda_{\theta}g^A_{\theta n}
\frac{1}{p^2_1
-m^2_{\theta}}\frac{1}{p^2_2-m^2}(-2\hat{p_1}p_1^{\mu}\gamma_5 +
 ....) \ee

where dots in r.h.s mean other kinematic structures (proportional
to $q$  e.t.c). From (6) one can easily obtain sum rule for axial
constant $g^A_{\theta n}$. For our sum rules we will use invariant
amplitude just at the kinematical structure $\hat{p_1}p_1^{\mu}$,
because, as it was discussed in $\cite{nblk}$, $\cite{nblk1}$,
$\cite{nblk2}$ the choice of the kinematic structures with maximal
number of momentum lead to better sum rules. So we obtain the
following sum rules

\be \lambda_n\lambda_{\theta}g^A_{\theta n}
e^{-(m_n^2/M_n^2+m_{\theta}^2/M_{\theta}^2)} = (-1/2)B_{\theta}B_n
\Pi^{QCD}\ee

where $B_{\theta}, B_n$ mean Borel transformation on pentaquark
and nucleon momenta correspondingly, and continuum extraction is
supposed.

By use of the equation of motions the eq.(4) close to the mass
shell can be rewritten \be \langle \theta^+ \mid j_{\mu}\mid n
\rangle =g^A_{\theta n} \bar{v}_n \Biggl (\gamma^{\mu}
+\frac{m_{\theta}+m_n}{q^2}q^{\mu}\Biggr )\gamma_5 v_{\theta} \ee

At the same time, the second term in  (4,8) correspond to the kaon
contribution to $\theta -n$ transition with lagrangian density
$L=ig_{\theta n k}v_{n}\gamma^5 v_{\theta} \phi_k$, so one can
write

\be \langle \theta^+ \mid j_{\mu}^5\mid n \rangle =g_{\theta n k}
\frac{q^{\mu}f_k}{q^2-m_k^2}\bar{v_{n}}\gamma^5 v_{\theta}\ee

Comparing (9) and  (8) one can found  (if we for a moment neglect
the kaon mass)

\be g_{\theta n k}f_k=(m_n+m_{\theta})g_{\theta n}^A\ee

This is the analog of the Golderberger-Trieman relation. Of course
the accuracy of this relation is about the scale of SU(3)
violation but as estimation of the value of $g_{\theta n k}$ it is
enough good. In \cite{ja1}, \cite{ja2} some general properties of
correlator (10) and correspondingly sum rules for $g^A_{\theta n}$
was obtained. First of all,  it was shown, that correlator (1)
vanishes in the chiral limit for any pentaquark current without
derivatives. That means, that first non-vanishing contribution to
sum rules give the operator with dimension $d=3$ (quark
condensate), so axial constant $g_{\theta n}^A$ should be
proportional to the quark condensate. An examples of corresponding
diagrams are shown on the  Fig.1a,b. But as was also shown in
these papers, diagrams like those on fig.1a  (i.e. without hard
gluon exchange) can not contribute to the sum rule. The reason is
that such diagrams, as one can easily check, are expressed in
terms of the following integrals

\be \int e^{i(p_1x-qy)} \frac{d^4 x d^4 y}{((x-y)^2)^n (x^2)^m}
\equiv \int \frac{e^{ip_1x}}{(x^2)^m} \frac{e^{-iq t}}{(t^2)^{n}}
d^4 xd^4t \ee

It is clear that such integrals have imaginary part on $p_2^2$ and
$q^2$ - the momentum of nucleon and axial current - but there is
no imaginary part on $p_1^2$ - the momentum of pentaquark. So we
come to the conclusion that such diagrams  correspond to the case,
when there is no $\Theta^+$ resonance in the pentaquark current
channel (this correspond to background of this decay). (Note, that
this conclusion don't depend on the fact that one ore more of the
quark propagators should be replaced by condensate, as we discuss
before). The double imaginary part on $p_1^2$, $p_2^2$  (i.e.
$\Theta^+$ resonance and baryon) appears only if one take into
account hard gluon exchange, and not arbitrary, but only those,
which connect the quark line, going to axial current vertex with
those going to an baryon vertex, (so that it provide the moment
exchange between these vertexes), as on Fig.1b. So we come to
conclusion, that if $\Theta^+$ is a genuine 5-quark state (not,
say, the $NK$ bound state), then the hard gluon exchange is
necessary, what leads to additional factor of $\alpha_s$. We see,
that pentaquark width $\Gamma_{\Theta} \sim \alpha^2_s \langle 0
\vert \bar{q} q \vert 0 \rangle^2$, i.e., $\Gamma_{\Theta}$ has
strong parametrical suppression.

\bigskip

{\bf \large Part 2. }

Let us now discuss the sum rules (7). First of all, of course, the
contribution of the operator of $d=3$ should be accounted. As we
discuss in previous section, only diagrams like those on Fig.1b
give the non-vanishing contribution to double dispersion relation.
It is necessary to note, that in the sense of discussion before
(see $\cite{ja2}$) it is quite necessary to keep only those part
of these diagrams, which have really imaginary part both on
$p_1^2$ and $p_2^2$ (i.e pentaquark and nucleon 4-momenta square).
This mean that double Borel transformation (for $p_1^2$ and
$p_2^2$ independently) in (7) is quite necessary.

 In what following, we will denote the result
of calculation (i.e. operator $d=3$ contribution to sum rules (7))
as $R_{d3}$

\be R_{d3}=1/(\lambda_n\lambda_{\theta})
e^{(m_n^2/M_n^2+m_{\theta}^2/M_{\theta}^2)}(-1/2)B_{\theta}B_n
\Pi^{QCD}\ee

It is clear, that

\be g^A_{\theta n}=R_{d3} +R_{d5} +...\ee

and if one confine itself only of $d=3$ operator contribution, (as
it was done in \cite{ja4})  then $g^A_{\theta n}=R_{d3}$. Let us
shortly discuss the results of \cite{ja4} (the higher order
operators contribution will be discussed in the next section).

First of all the calculation of the diagrams  Fig.1b is
technically enough complicated. One should pay special attention
to extract the terms, which have no imaginary part on pentaquark
4-momenta correctly. We perform calculation in the
x-representation in Euclid space, using standard exponential
representation of propagators and the relation $B
e^{-bp^2}=\delta(b-1/M^2)$.

We neglect in the calculations the effects of $s-quark$ mass,
which are very small.  We also use in calculation the fact, that
the ratio $A1=M_n^2/M_{\theta}^2$, (where $M_n^2, M_{\theta}^2$
are nucleon and pentaquark Borel masses) should be of order of
ratio of the corresponding mass square $m_n^2/m_{\theta}^2$, so we
can threats $A1$ as small parameter. But even at this
simplification the analytical answer is enormous large and is
expressed in terms of a very large number of different double (and
ordinary) integrals and it total size is very large (about some
hundred terms). That's why, unfortunately, we can not write down
pure analytical answer, and the last last stage  of calculation we
prefer to do only numerically . (Of course we check, that all
integrals converge if $Q^2 (=-q^2)$ is not equal to zero.)

Of course, as was discussed before, the result ($R_{d3}$) is
proportional to quark condensate and strong coupling constant,
i.e. $\alpha_s a$, where $ a = -(2 \pi)^2 \langle 0 \vert \bar{q}
q \vert 0 \rangle$. As a characteristic virtuality we chose the
Borel mass of nucleon, but one should note, that because $\alpha_s
a^2$ do not depend on normalization point, this choice is rather
unessential. We use the value of

\be \alpha_s a^2=0.23 Gev^6,~~
\bar{\lambda^2_n}=\lambda^2_n*32{\pi}^4=3.2Gev^6,~~
\bar{\lambda^2_{\theta}}=\lambda^2_{\theta}*(4{\pi})^8=12Gev^{12}\ee

(see \cite{bl2}, \cite{ja3}). To extract the continuum
contribution, usually one should write down explicitly double
dispersion integral, but it is too difficult to do technically in
our case, so we estimate continuum contribution by a usual factors
$E_0=1-e^{s_0/M^2}$ in both pentaquark and baryon channels. The
continuum dependence is not strong, we use standard value
$s_0=1.5GeV^2$ for nucleon and $s_0=4.-4.5GeV^2$ for pentaquark
current \cite{ja3}. We estimate the inaccuracy of the value of
$R_{d3}$ due to continuum contribution about $30\%$. In the paper
\cite{ja4}  only the contribution $R_{d3}$ to the value
$g^A_{\theta n}$ ( eq.(13)) was accounted and the estimation
$g^A_{\theta n} <1Mev$ was obtained. In this paper we
 also account the contribution of the higher dimension operators
 we are now going to discuss.

\bigskip

{\bf \large Part 3. }

There are large number of diagrams, corresponding to higher
dimension operators contribution (see some examples on figs.2,3,4
for dimension $d=5,7,9$ correspondingly). Numerically the main
contribution in our case give diagrams on Fig.4, (operator of the
dimension $d=9$), proportional to $a^3$, where  $a = -(2 \pi)^2
\langle 0 \vert \bar{q} q \vert 0 \rangle$). It is not surprising,
the same situation appear in many similar cases ( see \cite{nbl1}
for the case of nucleon, or \cite{ja3} for the case of
pentaquark). The reason is clear -each cut quark line lead, from
one side, to small factor -the quark condensate, but, from the
other side, it decreases the number of loops, so lead to
additional factor of order of $4{\pi}^2$, and this two factors
more or less compensate each other. That's why one can expect,
that main contribution of high dimension operator come from
diagrams on Fig. 4a-4b. When calculating this diagrams, one should
also carefully exclude those, which have no double imaginary part
both on $p_1^2$ and $p_2^2$. Examples of such diagrams which have
no double imaginary part are shown on Fig.5 - one can easily check
this just in the same way as was discussed in the previous chapter
(for diagrams on Fig.1a). So to estimate dimension $d=9$ operator
contribution $R_{d9}$ one should account only four types of
diagrams on Fig.4a,b (of course each of them with all possible
combinations). Fortunately, in this case it is possible to write
down analytical answer. If we neglect the small terms,
proportional to $s$-quark mass, and also for simplicity suppose
strange and light quark condensates to be equal, we can write:

\be
 R_{d9} = -0.209 (\bar{\lambda_n} \bar{\lambda_{\theta}})^{-1}
 e^{(m_n^2/M_n^2+m_{\theta}^2/M_{\theta}^2)}
 \frac{\alpha_s a^3}{\pi}
 \int\limits^{s_{\theta}}_{0} e^{-u/M^2_{\theta}} du
 \int\limits^{s_n}_{0} e^{-s/M^2_n} ds (\rho_1 +\rho_2/v) \ee

where $$\rho_1= (-3/8)u \delta(s)/Q^4$$

$$\rho_2= \frac{-1}{Q^2} + 2\frac{s^2 -u^2 +Q^4-1.5sQ^2}{Q^2v^2} +
7us\frac{u +Q^2-s}{v^4}$$

$v^2=(u+Q^2-s)^2 +4Q^2s$, and $\bar{\lambda_n},
\bar{\lambda_{\theta}}$  are defined in (14).

In the numerical analysis we suppose again $M_{\theta}^2=3M_n^2$,
as for $d=3$ operator case in previous section.

On Fig.6a,b the axial constant $g_{\theta n}^A =R_{d3}+R_{d9}$,
(see eq. (13)), obtained from sum rules, is shown for two values
of $Q^2$: $Q^2=1.5Gev^2$ (Fig.6a) and $Q^2=2.5Gev^2$ (fig.6b) as a
function of Borel mass of nucleon. Thick (upper) line mean the
total result for $g_{\theta n}^A$, thin line and dashed line -
$R_{d3}$ and $R_{d9}$ contributions correspondingly. One can see,
that $R_{d9}$ is smaller than $R_{d3}$ in this region and also the
Borel mass behavior of $g_{\theta n}^A$ is very good, so one can
suppose, that sum rule are reliable at this range of $Q^2$.

From the sense of sum rule, it is clear that we can found axial
constant $g_{\theta n}^A$ only at $Q^2$ not close to zero (about
$1 Gev^2$ or higher). And really, at $Q^2=1Gev^2$ $R_{d9}$ became
equal (and even more) than $R_{d3}$, so at this value sum rules
became very doubtful, and at lower values of $Q^2$ sum rules does
not exists.

 On the Fig.7 the $Q^2$
dependence of $g_{\theta n}^A$ is shown (at $M_n^2=1Gev^2$ and
$M_{\theta}^2=3M_n^2$ -thin (lower) line, and the same
for$M_n^2=1.2Gev^2$ - thick (upper) line ). One can see, that
they are practically identical.

Really we are interest the value $g_{\theta n}^A$ in the limit
$Q^2 \to 0$, which can't be calculated directly from S.R.,
obviously. But one can see from Fig.7, that $Q^2$ behavior is
found to be almost linear so one can extrapolate it to zero. We
found averaged (on Borel mass) $g_{\theta n}^A=0.034$ at $Q^2=0$.
Of course the accuracy of this value is not high, and highly
depend on the extrapolation (about a factor 1.5-2 for different
more or less reasonable extrapolations). Also large inaccuracy
appear due to:

a) the method itself has the accuracy of the order of the SU(3)
violation

c) accuracy in the value of $\lambda_{\theta}$ (about 20-30$\%$)

d) accuracy of sum rule approach, especially for pentaquark case
(see, for example, discussion in \cite{nar}).

For all this reason, we estimate  the value of axial constant
$g_{\theta n}^A(0)$  with inaccuracy about a factor two, so it can
varied from 0.017 to 0.068  with central point $g_{\theta
n}^A=0.034$. By use of eq (10) one can easily express the
pentaquark width in terms of $(g_{\theta n}^A)^2$, and we come to
conclusion, that our result for $\Gamma_{\Theta}$ can vary in the
region from 60$KeV$ to 1$MeV$. The central value $g_{\theta
n}^A=0.034$ correspond to width about 0.25$MeV$. Note, that our
result for width of the pentaquark (with positive parity) don't
contradict to value 0.75$MeV$, obtained in \cite{niels} also in
sum rules, but by quite differ method.

Our estimation of the pentaquark width  also is in the agreement
with the result \cite {dolg} (0.36$MeV$ with accuracy about
30$\%$), obtained from the ratio between numbers of resonant and
non-resonant charge exchange events.

Of course the accuracy of our result is not high, really we
predict only the order of magnitude of the width,  but it seems,
that more precise prediction for pentaquark width  in this
approach is impossible. That's why we don't discuss contribution
of operators $d=5,7$ e.t.c, because they are much smaller than
those we have accounted and their contribution is within of the
our accuracy.

The main conclusion is, that if the $\theta^+$ is genuine states
then sum rules really predict the very narrow width of pentaquark,
less than $1Mev$, (most probably about $0.25-0.5Mev$) and the
suppression of the width is both parametrical and numerical.

Author is thankful to B.L.Ioffe for useful discussions and
advises. This work was supported in part by CRDF grant
RUP2-2621-MO-04 and RFFI grant 06-02-16905.

\newpage

\begin{figure}
\epsfxsize=5cm \epsfbox{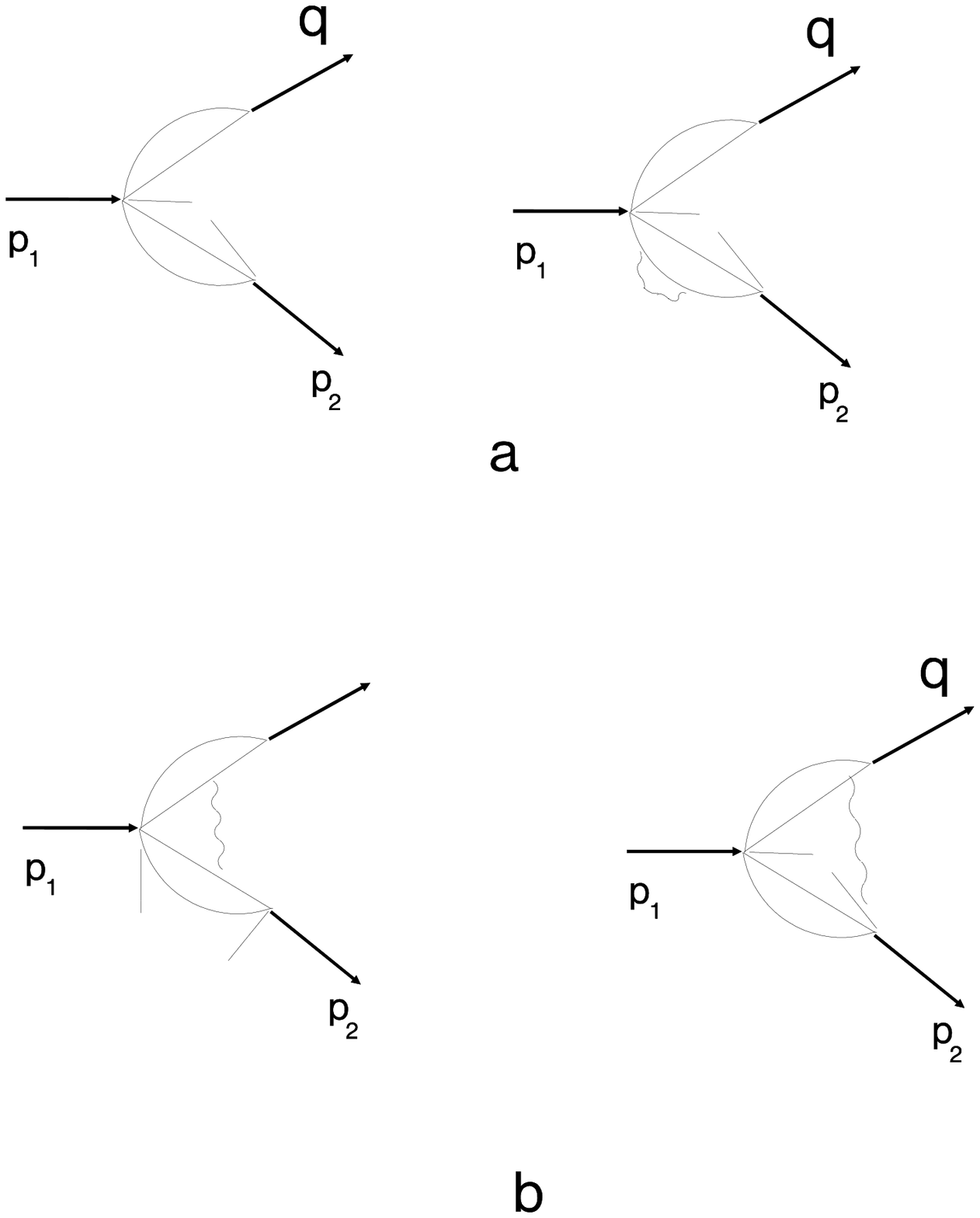} \caption{examples of the
diagrams for $d=3$ operator contribution}
\end{figure}

\begin{figure}
\epsfxsize=5cm \epsfbox{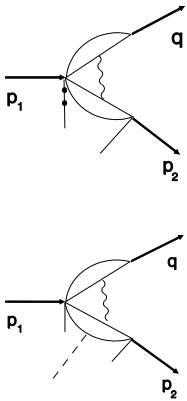} \caption{Examples of diagrams
for $d=5$ operator contribution, dashed line mean vacuum gluon
field, wave line -hard gluon exchange, circles mean derivatives}
\end{figure}

\begin{figure}
\epsfxsize=5cm \epsfbox{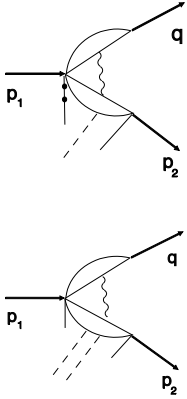} \caption{The same as in Fig. 2.
for $d=7$ operators}
\end{figure}

\begin{figure}
\epsfxsize=5cm \epsfbox{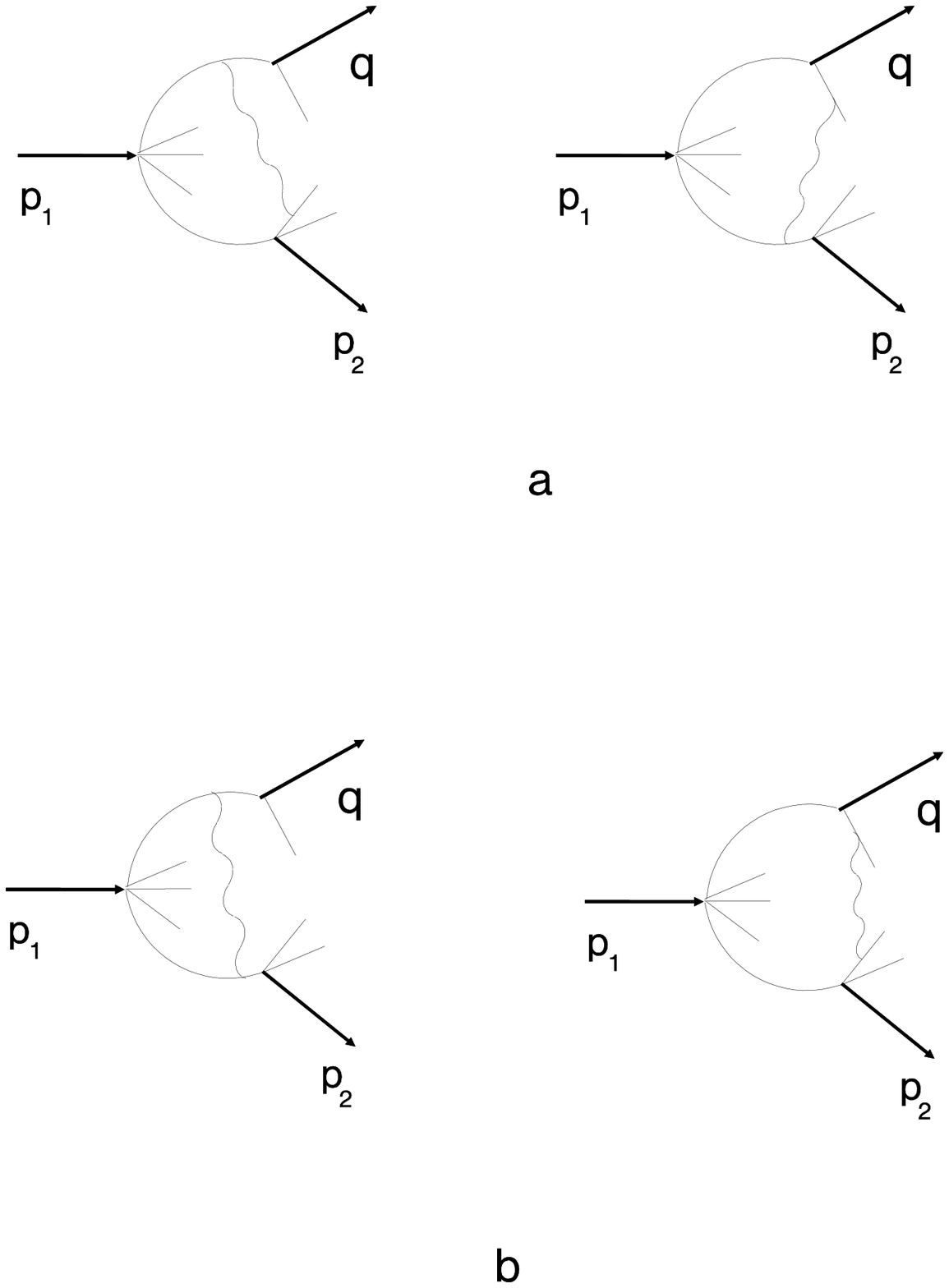} \caption{examples of the
diagrams for $d=9$ operator contribution}
\end{figure}

\begin{figure}
\epsfxsize=5cm \epsfbox{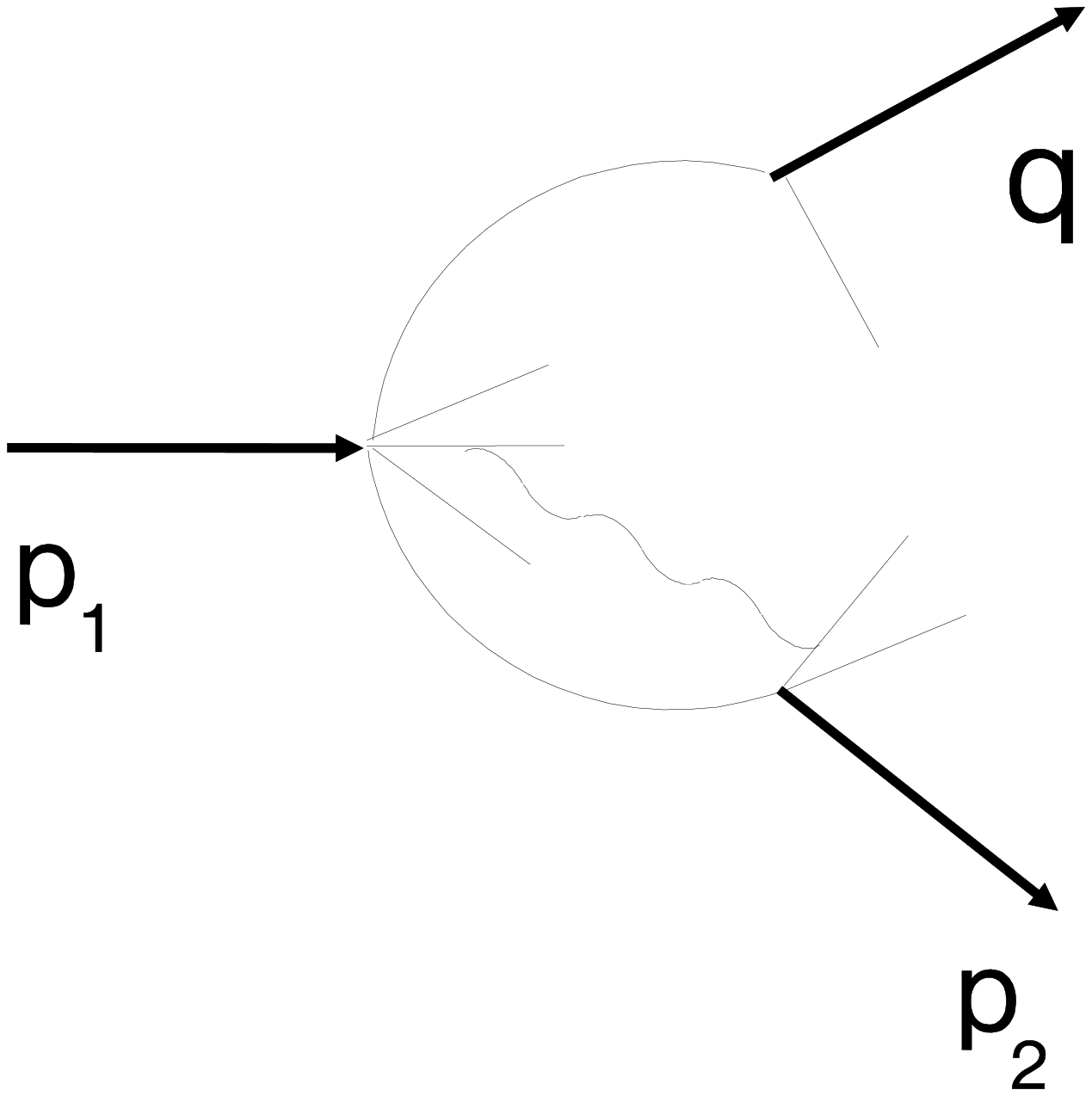} \caption{}
\end{figure}

\begin{figure} \epsfxsize=15cm \epsfbox{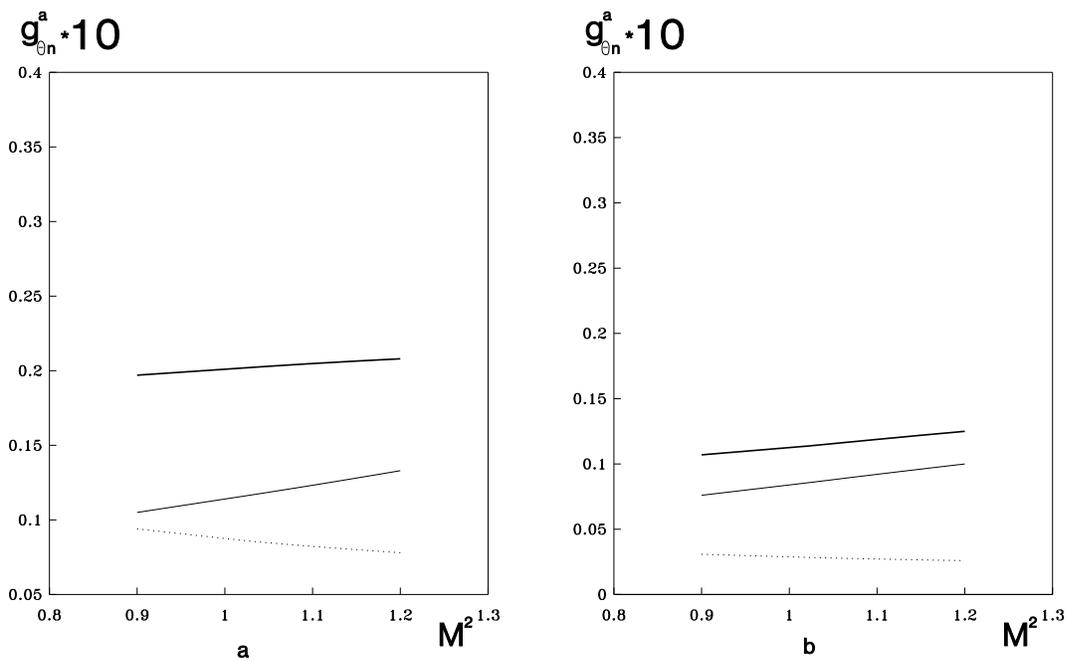}
\caption{$g^A_{\theta n}$ dependence on Borel mass for: a)
$Q^2=1.5GeV^2$, b) for $Q^2=2.5GeV^2$}
\end{figure}

\begin{figure}
\epsfxsize=5cm \epsfbox{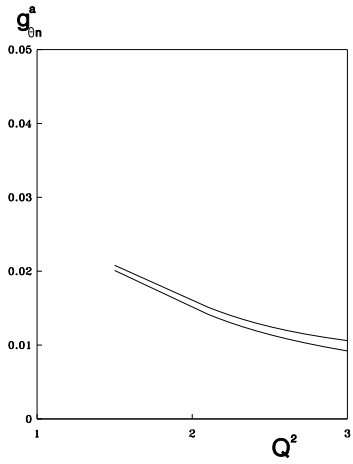} \caption{ $Q^2$ dependence of
$g^A_{\theta n}$ for$M_n^2=1.2GeV^2$ -thick (upper)
 line and for $M_n^2=1.GeV^2$ - thin (lower) line}
\end{figure}

\end{document}